\documentclass[twocolumn,a4paper,10pt]{quantumarticle}
\pdfoutput=1
\usepackage{amsmath,amssymb,amsfonts}
\usepackage{graphicx}
\usepackage{adjustbox}
\usepackage[english]{babel}
\usepackage{algpseudocode}
\usepackage{multirow}
\usepackage{booktabs}
\usepackage{array}
\usepackage{calc}
\usepackage{tabularx}
\usepackage{url}
\usepackage{placeins}   

\newcounter{myalgorithm}
\renewcommand{\themyalgorithm}{\arabic{myalgorithm}}
\usepackage[numbers,sort&compress]{natbib}
\usepackage{hyperref}
\begin{document}

\author{Jacobo Padín-Martínez}
\affiliation{International Quantum Center - FSAS Technologies (Fujitsu)}
\author{Vicente P. Soloviev}
\affiliation{Fujitsu Research of Europe Ltd.}
\author{Alejandro Borrallo-Rentero}
\author{Antón Rodríguez-Otero}
\author{Raquel Alfonso-Rodríguez}
\affiliation{International Quantum Center - FSAS Technologies (Fujitsu)}
\author{Michal Krompiec}
\affiliation{Fujitsu Research of Europe Ltd.}

\title{Progressive Binarization - Pauli Correlation Encoding:\newline a Continuation Method for Constrained Optimization}

\begin{abstract}
Pauli Correlation Encoding (PCE) reduces the qubit requirements of quantum optimization by embedding 
the problem variables into the expectation values of Pauli observables, so that the number of qubits can 
be much smaller than the number of variables. PCE has not yet been studied for constrained optimization. 
We extend it to constrained combinatorial problems, using the budget-constrained MinCut as a case study, 
and show that the standard formulation fails to reliably enforce the constraint: feasibility hinges on the 
binarization of the encoded variables, which depends sensitively on hyperparameters that are hard to tune 
and do not transfer across instances. To address this, we introduce Progressive Binarization-PCE (PB-PCE), 
an adaptive continuation scheme that progressively increases the binarization parameter while re-optimizing 
the circuit from the previous solution, driving the variables towards the binary domain. PB-PCE attains 
near-complete constraint satisfaction (88--100\%) and smaller cut sizes than standard PCE, with a number 
of stages (10--20) essentially independent of problem size, solving instances of up to 300 variables with up to 9 
simulated qubit circuits.
\end{abstract}

    \maketitle

    \section{Introduction}
        Quantum optimization has emerged as a prominent area of research within quantum computing over recent decades \citep{abbas_challenges_2024}. The steady increase in the number of qubits available in quantum processing units (QPUs) has made it possible to address and solve problem instances that more closely resemble real-world use cases, bridging the gap between classical computational approaches and industrial applications.
        
        Among the most established approaches, variational quantum algorithms (VQAs) leverage tunable quantum circuits optimized through classical feedback loops, making them particularly suitable for near-term noisy devices \citep{cerezo_variational_2021}. An example of such algorithms is the Quantum Approximate Optimization Algorithm (QAOA) \citep{farhi_quantum_2014}, designed to approximate solutions to NP-hard graph-based problems. Other examples of VQAs include the Variational Quantum Eigensolver (VQE) \citep{peruzzo_variational_2014}, traditionally used for quantum chemistry problems. Quantum annealing (QA) \citep{falco_introduction_2011} represents another influential paradigm, relying on adiabatic evolution to guide a quantum system toward the ground state of a problem Hamiltonian; this method has been widely explored in both academic research and industry \citep{yarkoni_quantum_2022}. 
        
        Most of the approaches discussed above share a common characteristic: the mapping from classical optimization variables to qubits is typically performed through a one-hot encoding, whereby each variable is assigned to a qubit in a one-to-one manner. This strategy becomes a significant limitation when scaling to realistic scenarios involving a large number of variables. Recently, the Pauli Correlation Encoding (PCE) \citep{sciorilli_towards_2025} algorithm was proposed to overcome this limitation by allowing the number of qubits in the system to be substantially smaller than the number of optimization variables. This is achieved by encoding variable information into expectation values of Pauli observables rather than directly into the quantum state of individual qubits.
        
        Given its recent introduction, the literature on PCE remains limited. Nevertheless, the approach has already been applied to realistic financial scenarios, such as portfolio optimization \citep{soloviev_large-scale_2025}, and its adoption is expected to grow, as major platforms are dedicating specific tutorials and resources to the method \citep{noauthor_pauli_nodate}. Despite this growing interest, systematic studies of PCE remain scarce, making further theoretical and empirical insights particularly valuable.
        
        To the best of our knowledge, constrained optimization problems have not yet been analyzed within the PCE framework. In this work we extend PCE to constrained combinatorial optimization, using the budget-constrained MinCut as a representative case study, and we show that the standard formulation does not reliably enforce the constraint. We trace this failure to the binarization of the encoded variables: feasibility requires the relaxed variables to attain binary values, but the degree of binarization depends sensitively on hyperparameters that are hard to tune and do not transfer across instances. Building on this analysis, we introduce \emph{Progressive Binarization-PCE} (PB-PCE), which we formulate as an adaptive continuation (graduated-optimization) scheme over the binarization parameter: 
        starting from a smooth relaxation of the sign function, we converge
        to the optimal solution by driving the variables to the binary domain.

        Our main contributions are the following:
        \begin{itemize}
            \item \textbf{Constrained optimization with PCE.} We extend PCE to constrained combinatorial optimization and show that the standard formulation fails to reliably satisfy the constraint, identifying the binarization of the encoded variables as the governing factor.
            \item \textbf{Hyperparameter analysis.} We systematically analyze the PCE's binarization parameter $\alpha$ and the constraint's penalty parameter $\beta$, showing that they are hard to tune and do not transfer across instances. We propose a graph-derived penalty that removes the per-instance tuning of $\beta$, showing that the main governance is given by the binarization parameter $\alpha$.
            \item \textbf{PB-PCE.} We introduce Progressive Binarization-PCE, an adaptive continuation scheme over the binarization parameter, which attains near-complete constraint satisfaction ($88\text{--}100\%$) and consistently smaller cut sizes than standard PCE, with a number of stages ($10\text{--}20$) essentially independent of the problem size.
            \item \textbf{Scalability.} Using PB-PCE we solve constrained instances of up to $300$ variables with only $9$-qubit circuits.
        \end{itemize}
        All results are obtained in noiseless statevector simulation, using a single-layer brickwork ansatz, with simulated annealing as the classical reference for normalized cut sizes.
        
        The outline of the paper is as follows: Section~\ref{sec_pce} describes the PCE approach in detail; Section~\ref{sec_constrained_problems} presents the specific problem used in the experimentation; Section~\ref{sec_analysis} provides the results and analysis of the PCE baseline; Section~\ref{sec_iterative_PCE} introduces the proposed Progressive Binarization-PCE (PB-PCE) method and additional benchmarking results; finally, Section~\ref{sec_conclusions} concludes the paper with further considerations and directions for future research.

    \section{Pauli Correlation Encoding}
    \label{sec_pce}
        \begin{figure*}[!ht]
            \includegraphics[width=\textwidth]{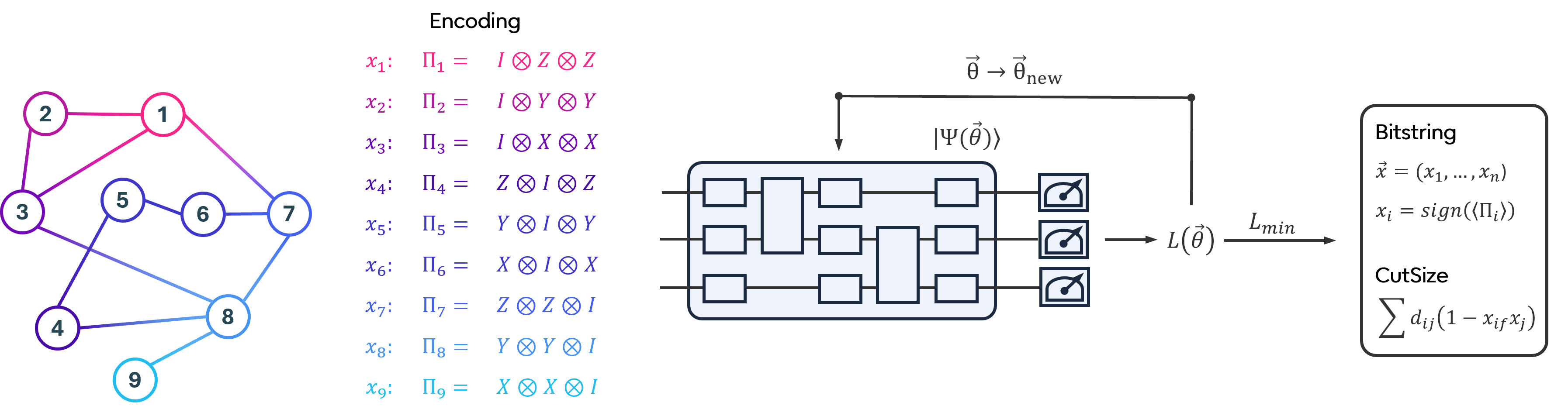}
            \caption{Pauli Correlation Encoding (PCE) optimzation scheme.}
            \label{fig: pce optimization scheme}
        \end{figure*}

        The Pauli Correlation Encoding (PCE) is a quantum algorithm designed to solve combinatorial problems (originally introduced for QUBO problems) on gate-based quantum computers. Unlike approaches such as QAOA, which encode binary variables in the computational basis states of qubits, PCE encodes the variables in the signs of the expectation values of a predefined set of Pauli string operators. These operators are constructed as tensor products of Pauli matrices, with each string containing only a single type of Pauli matrix:
        \begin{equation}
            \Pi_{i} \ \varepsilon \left\{X^{\otimes k}\otimes\mathbb{I}^{\otimes m-k},Y^{\otimes k}\otimes\mathbb{I}^{\otimes m-k},Z^{\otimes k}\otimes\mathbb{I}^{\otimes m-k}\right\}
        \end{equation}
        or any of their permutations, where $k$ is the order (number of Pauli matrices per string), and $m$ is the number of qubits. For example, in a three-qubit system with $k=2$, the possible operators include $\{$IXX, IYY, IZZ, XIX, YIY, ZIZ, XXI, YYI, ZZI$\}$, effectively encoding 9 variables using only 3 qubits.
        
        In the original PCE paper \citep{sciorilli_towards_2025}, the authors applied this approach to solve Max-Cut problems, formulated as the minimization of the following QUBO:
        \begin{equation}
        \mathrm{MaxCut} = \sum_{i>j}^{n}\sum_{j=0}^{n-1} d_{ij}\left(1-x_i x_j\right),
        \end{equation}
        where $d_{ij}$ denotes the weight of the edge that connects nodes $i$ and $j$, and $x_i\in\{0,1\}$ represents the binary variable associated with node $i$, indicating which side of the cut the node is assigned. The objective function corresponds to the cut value, defined as the sum of the weights of the edges connecting nodes belonging to different partitions. For instance, if $x_i = 0$ and $x_j = 1$, nodes $i$ and $j$ are placed on opposite sides of the cut and the corresponding edge contributes to the objective function. In the PCE algorithm, binary variables are substituted by the sign function, i.e. $x_i=\mathrm{sign}\left(\left\langle\Pi_i\right\rangle\right)$:
        \begin{equation}
                \mathrm{MaxCut}=  \sum_{i>j}^{n}\sum_{j=0}^{n-1} {d_{ij}\left[1-\mathrm{sign}(\langle\Pi_i\rangle) \ \mathrm{sign}(\langle\Pi_j\rangle)\right]}
        \end{equation}
        However, in practice, the sign function is replaced by the hyperbolic tangent, which provides a smooth approximation better suited for classical optimizers \citep{noauthor_variational_nodate}. The loss is therefore written as a function of the circuit parameters $\vec{\theta}$ and of a sharpness parameter $\alpha$:
        \begin{multline}
            L_\alpha(\vec{\theta})= \sum_{i>j}^{n}\sum_{j=0}^{n-1} {d_{ij}\left[1-\tanh{\left(\alpha\left\langle\Pi_i\right\rangle\right)}\tanh{\left(\alpha\left\langle\Pi_j\right\rangle\right)}\right]}
            \label{eq: pce loss}
        \end{multline}
        where the expectation values $\langle\Pi_i\rangle = \langle\Psi(\vec{\theta})|\Pi_i|\Psi(\vec{\theta})\rangle$ depend on the circuit parameters, and $\alpha$ controls how sharply the smooth proxy $\tanh(\alpha\langle\Pi_i\rangle)$ approximates the sign function; we therefore refer to $\alpha$ as the \emph{binarization parameter}. Usually, a regularization term $L^{reg}$ is introduced in this expression, but it will be discussed later. 
        
        As $\alpha\to\infty$ the proxy recovers the sign function, so that $L_\alpha(\vec{\theta})$ interpolates between a smooth relaxation and the original objective, a property that is central to the method introduced in Section~\ref{sec_iterative_PCE}. The optimization workflow, detailed in Figure~\ref{fig: pce optimization scheme}, is as follows: 
        \begin{enumerate}
            \item Prepare a parametrized quantum state $|\Psi(\vec{\theta})\rangle$.
            \item Compute the expectation values $\langle\Pi_i\rangle$ to evaluate the loss function $L_\alpha(\vec{\theta})$.
            \item Use a classical optimizer to update the parameters for the next iteration.
            \item Repeat the procedure until the loss is minimized.
            \item Once convergence is reached , construct the bitstring solution:\\$x=\left(x_1,x_2,\ldots,x_n\right)$ with $x_i=\mathrm{sign}\left(\left\langle\Pi_i\right\rangle\right)$. 
            \item Compute the cut size value using the bitstring solution in the original QUBO expression Eq.~(\ref{eq: mincut}).
        \end{enumerate}

    \section{Budget-Constrained MinCut Problem}
    \label{sec_constrained_problems}
    
        Many real-world optimization problems involve additional constraints that go beyond the standard unconstrained MinCut or MaxCut formulations. In particular, budget or balance constraints naturally arise in applications where the two partitions induced by the cut must satisfy predefined size, capacity, or cost requirements \citep{puerto_budget-constrained_2023, engelberg_cut_2007, chekuri_maximum_2004, takei_optimal_2015}. Examples include resource allocation \citep{shi_resource_2010}, and clustering problems \citep{rose_constrained_1993}. In this study we will focus on the mincut problem with a budget constraint term.
        
        First, a MinCut problem can be formulated as minimizing the following QUBO problem:
        \begin{equation}
            \mathrm{MinCut}=\sum_{i=1}^{n-1}\sum_{j=i+1}^{n}{d_{ij}\left(x_i-x_j\right)^2} 
            \label{eq: mincut}
        \end{equation}
        A budget-constrained MinCut is formulated by introducing a constraint term as follows:
        \begin{equation}
            \mathrm{MinCut}=\sum_{i=1}^{n-1}\sum_{j=i+1}^{n}{d_{ij}\left(x_i-x_j\right)^2}+\beta\left(\sum_{i=1}^{n}x_i-c\right)^2
            \label{eq: mincut with constraint}
        \end{equation}
        where $c$ denote the number of nodes assigned to one of the subgroups. For example, the divisions $1:7$, $2:6$, $3:5$ and $4:4$ 
        can be obtained by setting $c=1,2,3,4$ (or equivalently $7,6,5,4$). In general, for a system with n nodes, $c\in\left[1,\ n/2\right]$, 
        since the values $c > N/2$  reproduce the same divisions, i.e., a division $c:N-c$ is equivalent to $N-c:c$.

        To solve the budget-constrained MinCut QUBO using the PCE, first it has to be expressed in terms of binary variables $z_i=\{-1,1\}$. Let $x_i=\{0,1\}$, the change of variables is defined as:
        \begin{equation}
            x_i=\frac{z_i+1}{2}
        \end{equation}
        the cut value term can be rewritten as:
        \begin{multline}
            \sum_{i=1}^{n-1}\sum_{j=i+1}^{n}{\frac{1}{4}d_{ij}\left(z_i-z_j\right)^2}\ =
            \\
            =\sum_{i=1}^{n-1}{\sum_{j=i+1}^{n}{\frac{1}{4}d_{ij}\left(-2z_iz_j+z_i^2+z_j^2\right)}\ }
        \end{multline}
        since $z_i^2=1$, this simplifies to:
        \begin{equation}
            \sum_{i=1}^{n-1}{\sum_{j=i+1}^{n}{\frac{1}{2}d_{ij}\left(1-z_iz_j\right)}\ }
        \end{equation}
        The constraint term, applying the same change of variables yields:
        \begin{multline}
            \beta\left(\sum_{i}^{n}x_i-c\right)^2=\beta\left[\frac{1}{2}\left(\sum_{i}^{n}{z_i+1}\right)-c\right]^2=
            \\
            =\beta\left[\frac{1}{2}\left(\sum_{i}^{n}{z_i+1}-2c\right)\right]^2
        \end{multline}
        ignoring the constant factor 1/2, which does not affect the optimization process:
        \begin{equation}
            \beta\left[\left(\sum_{i}^{n}z_i\right)+n-2c\right]^2
        \end{equation}
        the last step is to apply the substitution $z_i \to \tanh\!\left(\alpha\langle\Pi_i\rangle\right)$, obtaining the constrained loss as a function of $\vec{\theta}$ and $\alpha$:
        \begin{multline}
            L_\alpha(\vec{\theta}) = \sum_{i=1}^{n-1}\sum_{j>i}^{n}
            \frac{1}{2}\!d_{ij}\!\left[1-\tanh\!\left(\alpha\langle\Pi_i\rangle\right)
            \tanh\!\left(\alpha\langle\Pi_j\rangle\right)\right] + \\
            + \beta\left[\sum_{i}^{n}\tanh\!\left(\alpha\langle\Pi_i\rangle\right)
            -\left(n-2c\right)\right]^2
            \label{eq: pce mincut with constraint}
        \end{multline}
        The regularization term is not included in this expression.

        The fundamental challenge in applying Pauli Correlation Encoding (PCE) to constrained cut problems stems from its inherently non-binary nature. Algorithms such as the Quantum Approximate Optimization Algorithm (QAOA) are intrinsically tailored to binary optimization, as their solutions are obtained through qubit measurements that naturally yield binary outcomes in the computational basis, resulting in a bitstring representation of the solution. In an ideal setting, PCE would exhibit an analogous behavior with the sign function, however, in practice, the sign function is relaxed to $\tanh({\cdot})$, allowing the encoded variables to take continuous values in the interval $\left[-1,1\right]$ rather than being restricted to the binary set $\{-1,1\}$.
        
        This relaxation gives rise to two critical consequences. First, the cut-value term in the loss function is minimized over a continuous domain. As a result, configurations of the variables with decimal values can yield lower loss values than any valid binary configuration, potentially leading to incorrect cut assignments once the solution is discretized. Second, the constraint term may fail to be satisfied: the optimizer can find a configuration that minimizes the penalty over the real domain while not satisfying the proper integer condition. In fact, for a fixed value of $c$, the constraint defined in Eq.~\ref{eq: pce mincut with constraint} is fulfilled when the sum of the $\tanh{(\alpha\langle\Pi_i\rangle)}$ variables equals $n - 2c$. In the binary case, where $\tanh{(\alpha\langle\Pi_i\rangle)} \in \{-1,1\}$, this condition guarantees that exactly $c$ nodes are assigned to one side of the cut. In the continuous regime, however, this equality is generally not enforced: instead, the optimizer seeks a real-valued configuration that minimizes the constraint penalty, which does not necessarily correspond to selecting precisely $c$ nodes.
               
        In this paper we will analyze and address this problematic, by using the PCE on the budget-constrained MinCut problems, where the cut must satisfy a global constraint on the number of nodes assigned to each partition.

    \section{PCE Analysis}
    \label{sec_analysis}

    In this section, we analyze the PCE algorithm in the context of the budget-constrained MinCut optimization problem described in Section~\ref{sec_constrained_problems}. We focus on how each component of the algorithm influences the solution, with particular emphasis on constraint satisfaction and its relationship with the final values of the variables. The graph instances used throughout this section are summarized in Table~\ref{fig_small_graphs_description}. The ansatz employed in this work is the same as that used in the original PCE study, the Brickwork Ansatz, and only a single layer is used.
        \begin{figure*}[t]
            \centering
            \includegraphics[width=\textwidth]{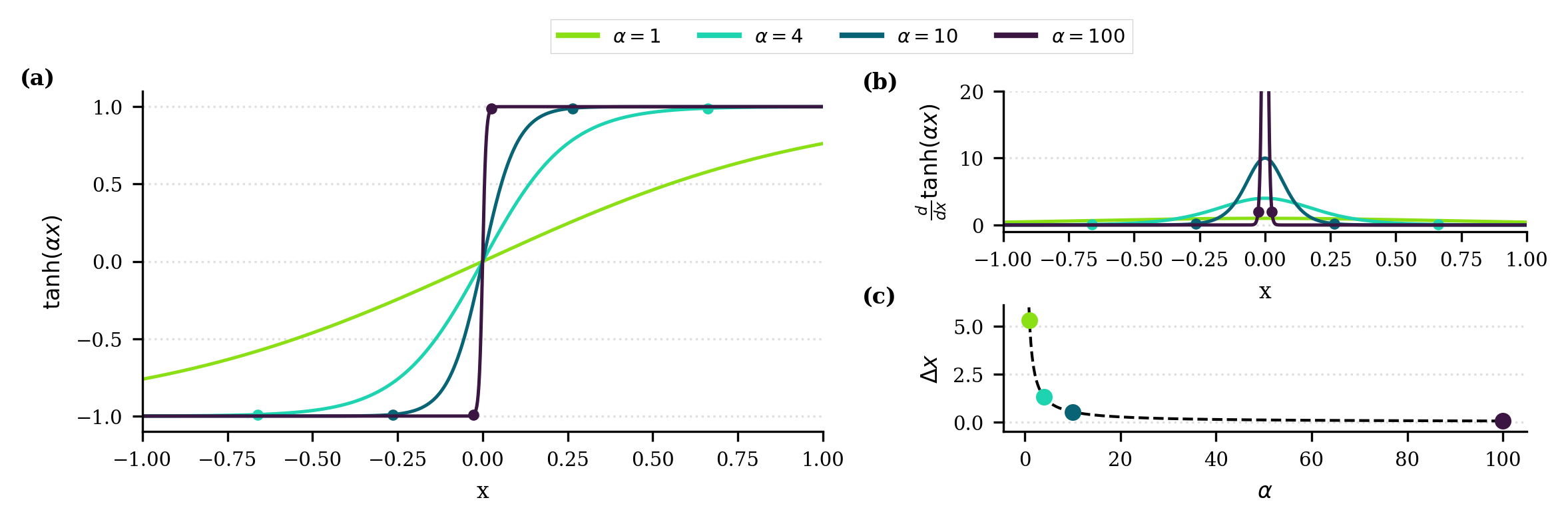}
            \caption{(a) $\tanh\left(\alpha\left\langle\mathrm{\Pi}\right\rangle\right)$ function. (b) $\tanh\left(\alpha\left\langle\mathrm{\Pi}\right\rangle\right)$ derivative. In both plots a point 
            is marked where $\tanh\left(\alpha\left\langle\mathrm{\Pi}_i\right\rangle\right)=\pm0.99=x_{\pm0.99}$. (c) Width of the plateau region $\mathrm{\Delta x}=x_{+0.99}-x_{-0.99}$ vs $\alpha$ is shown. The points highlighted 
            are for those $\alpha$ in the plots}
            \label{fig: tanh and derivate}
        \end{figure*}
        For the analysis of the results, we introduce the following metrics:
        \begin{enumerate}
            \item \textbf{Constraint success ratio}: Denoted as $\mathbf{\varepsilon}_{c}$, this value represents the efficiency with which the algorithm fulfills the constraint. Given $N$ executions of the PCE, being $N_c$ the number of executions in which the cut satisfied the constraint, this ratio is:
            \begin{equation}
                \varepsilon_{c} = \frac{N_c}{N}
            \end{equation}
            If $\varepsilon_{c}=0$ there was not any execution which satisfied the constraint, and therefore there is not any valid solution. If $\varepsilon_{c}=1$ all executions satisfied the constraint and every solution is valid.
            \item \textbf{Binarization}: The ratio of variables that effectively attain binary values after optimization. We consider that a variable 
            $\tanh{(\alpha\langle\Pi_i\rangle)}$ is binarized if $|\tanh{(\alpha\langle\Pi_i\rangle)}|>0.9$. Defining $V$ as the set of binarized variables: \[V = \left\{\, i \;\middle|\; \left| \tanh\!\bigl(\alpha \langle \Pi_i \rangle\bigr) \right| > 0.9 \,\right\}\]the binarization is computed as:
            \begin{equation}
                \mathrm{Binarization}=\frac{ |V|}{n}
            \end{equation}
            If $\mathrm{Binarization=0}$ means no variable was binarized and all remained in the real regime between $[-0.9,0.9]$. On the contrary, if 
            $\mathrm{Binarization}=1$ all the variables were binarized. 
            \item \textbf{CutSize}: The cut size value given by the bitstring solution $z=\left(z_1,z_2,\ldots\ z_{n}\right)$:
            \begin{equation}
                \mathrm{Cut Size} =\sum_{i=1}^{n-1}{\sum_{j=i+1}^{n}{\frac{1}{2}d_{ij}\left(1-z_iz_j\right)}\ }
            \end{equation}
            To facilitate comparisons across different graph instances and values of the constraint parameter $c$, all CutSize values are reported normalized with respect to the solution obtained for the same graph and constraint using simulated annealing~\cite{van_laarhoven_simulated_1987}. This reference solution provides a high-quality baseline against which the performance of the PCE algorithm can be evaluated.

        \end{enumerate}

        In all the results presented in this work, the CutSize is computed only for those simulations in which the constraint is satisfied. This restriction ensures that these metrics reflect the quality of valid solutions exclusively. Since both quantities are intended to evaluate the performance of the algorithm in producing meaningful cuts, it would be inconsistent to include cases where the constraint is violated, as such configurations do not represent feasible solutions to the problem. 
        \setlength{\tabcolsep}{4pt}  
        \begin{table}[htbp]
        \centering
        \begin{tabular}{lrrrc}
        \toprule
        Nodes & Edges & Strength & Strength Std & Connected\\
        \midrule
        6  & 15  & 31.81 & 6.68 & \checkmark \\
        14 & 91  & 14.04 & 12.23 & \checkmark \\
        18 & 153 & 12.81 & 9.60 & \checkmark \\
        20 & 190 & 30.46 & 9.72 & \checkmark \\
        25 & 300 & 22.00 & 5.91 & \checkmark \\
        \bottomrule
        \end{tabular}
        \caption{Key structural properties of the analyzed graphs. Mean Strength denotes the average node strength, computed as the mean incident edge weight per node. Strength Std denotes the average standard deviation of incident edge weights across nodes. The final column indicates whether each graph forms a single connected component.}
        \label{fig_small_graphs_description}
        \end{table}
        \label{sec_results}
            \subsection{PCE's binarization parameter $\alpha$}
                \begin{figure*}[t]
                    \centering
                    \includegraphics[width=\textwidth]{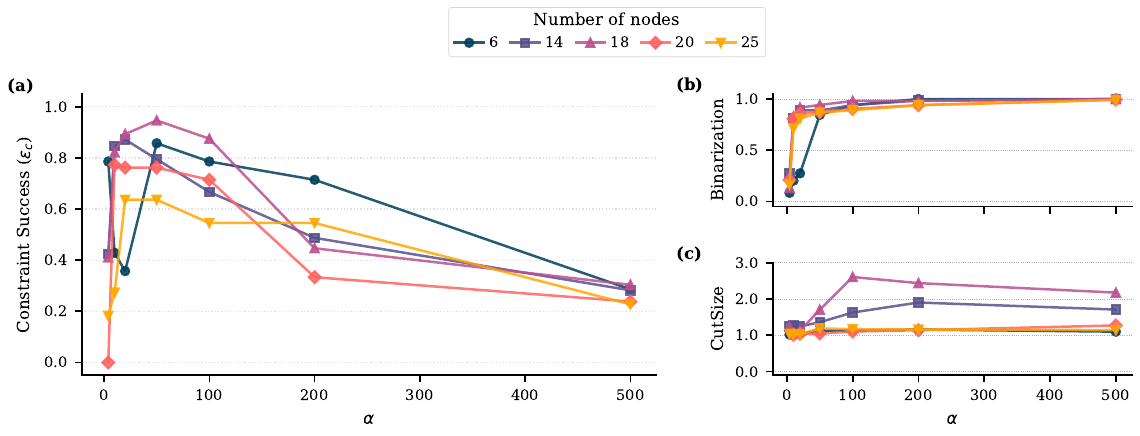}
                    \caption{(a) Constraint success ratio $\varepsilon_c$ as a function of $\alpha$.
                    (b) Binarization as a function of $\alpha$.
                    (c) CutSize as a function of $\alpha$.
                    Each graph was executed 10 times for all values of the constraint parameter $c \in [2, m/2]$, where $m$ denotes the number of nodes. Reported values correspond to averages over all runs. The classical optimizer used was SLSQP. The penalty parameter was fixed to $\beta = 1000$.}
                    \label{fig: alpha performance with SLSQP}

                    \vspace{\floatsep}

                    \includegraphics[width=\textwidth]{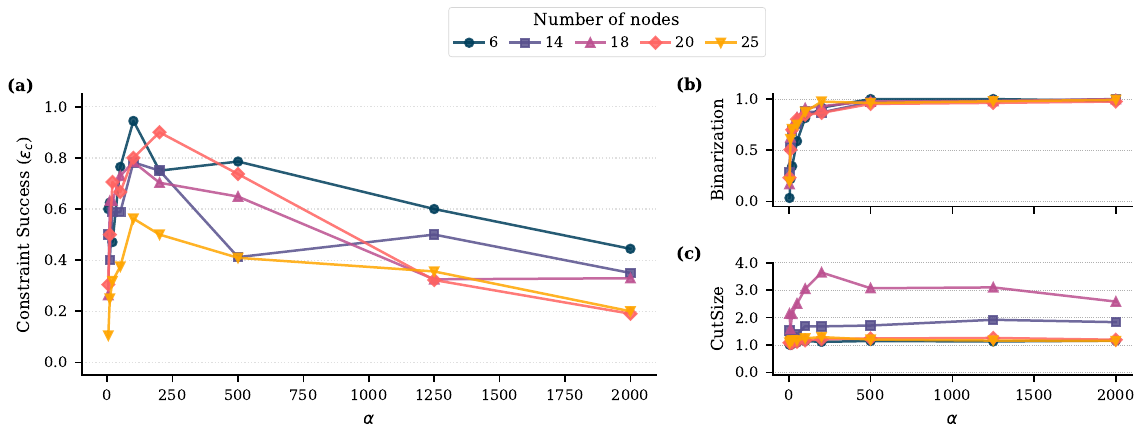}
                    \caption{(a) Constraint success ratio $\varepsilon_c$ as a function of $\alpha$.
                    (b) Binarization as a function of $\alpha$.
                    (c) CutSize as a function of $\alpha$.
                    Each graph was executed 10 times for all values of the constraint parameter $c \in [2, m/2]$, where $m$ denotes the number of nodes. Reported values correspond to averages over all runs. The classical optimizer used was Nelder-Mead. The penalty parameter was fixed to $\beta = 1000$.}
                    \label{fig: alpha performance with nelder mead}
                \end{figure*}
                The parameter $\alpha$ controls the smoothness of the effective sign function. Since the expectation values are bounded within the interval $\left[-1,1\right]$, the parameter $\alpha$ must be sufficiently large for the hyperbolic tangent to approach its asymptotic values of -1 and 1 within this interval. For instance, using $\alpha=1$ precludes reaching this limits, resulting in insufficient binarization (see Figure~\ref{fig: tanh and derivate}).
                We evaluated the PCE algorithm over a wide range of values $\alpha \in [4, 1000]$. Figure~\ref{fig: alpha performance with SLSQP} illustrates the impact of $\alpha$ on both the constraint success ratio $\varepsilon_c$ and the binarization. For low values of $\alpha$, both $\varepsilon_c$ and the binarization remain low, increasing progressively as $\alpha$ grows. The maximum value of $\varepsilon_c$ is achieved in the same range where the binarization approaches unity. Beyond this regime, larger values of $\alpha$ lead to fully binarized variables; however, the constraint success ratio is drastically lower. In addition, the CutSize is observed to increase with $\alpha$, particularly for the 14-node and 18-node graph instances, indicating that the algorithm struggles to identify better minima.
                This behavior reflects an intrinsic limitation of approximating the sign function. Beyond a certain value of $\alpha$, the optimizer fails to identify parameter configurations that further minimize the constraint term. Moreover, even in cases where the constraint is satisfied, the algorithm frequently converges to suboptimal solutions. 

                The optimizer employed in our experiments is the same as that used in the original PCE work, the SLSQP algorithm. This optimizer determines the optimization trajectory by exploiting gradient information of the loss function. In the PCE, the loss depends explicitly on the derivatives of the $\tanh(\cdot)$ function. As $\alpha$ increases, the derivative rapidly vanishes over a broad region of the variable domain, as shown in Figure~\ref{fig: tanh and derivate}(b). 
                
                To further investigate whether this behavior is related to the use of a gradient-dependent optimizer such as SLSQP, we repeated the same analysis using a gradient-free method, the Nelder-Mead algorithm~\citep{nelder_simplex_1965}. The corresponding results are shown in Figure~\ref{fig: alpha performance with nelder mead}. A qualitatively similar behavior is observed when using Nelder-Mead, indicating that the observed effect is independent of the use of gradient information.
                Consequently, the limitation cannot be attributed to the gradients evaluating to zero. Instead, it arises from the structure of the loss landscape itself. As $\alpha$ increases, the effective region of the parameter space that meaningfully influences the loss function becomes increasingly narrow. This effect is illustrated in Figure~\ref{fig: tanh and derivate}(c), where the width of the plateau associated with the $\tanh$ function is shown as a function of $\alpha$. As $\alpha$ grows, this width shrinks toward zero, severely restricting the region in which parameter variations lead to appreciable changes in the loss. As a result, both gradient-based and gradient-free optimizers are unable to efficiently explore the landscape, leading to premature convergence to suboptimal solutions.

                Therefore, the optimal value of the parameter $\alpha$ must be sufficiently large to ensure that the variables are properly binarized and that the resulting solution satisfies the constraint, while at the same time avoiding stalling of optimization in flat regions of the landscape. The original PCE paper suggests that the parameter $\alpha$ scales as $\alpha \sim n^{k/2}$, where $n$ denotes the number of qubits and $k$ the number of Pauli operators per Pauli string. In this experiments we have used $k=2$. Since the instances studied here are encoded in only 3 to 5 qubits, this prescription yields $\alpha \sim n \sim 4$, which coincides with the leftmost point of Figures~\ref{fig: alpha performance with SLSQP} and~\ref{fig: alpha performance with nelder mead}. At that value the constraint success ratio is clearly worse than the one attained around $\alpha \sim 100$, showing that the scaling suggested for the unconstrained problem is not an adequate choice once a constraint has to be enforced.
                In summary, $\alpha$ constitutes a critical hyperparameter in the PCE algorithm. While strong binarization seems to be a necessary condition for satisfying the constraint, it is not sufficient.
            \subsection{Penalty parameter $\beta$}
                \begin{figure*}[t]
                    \includegraphics[width=\linewidth]{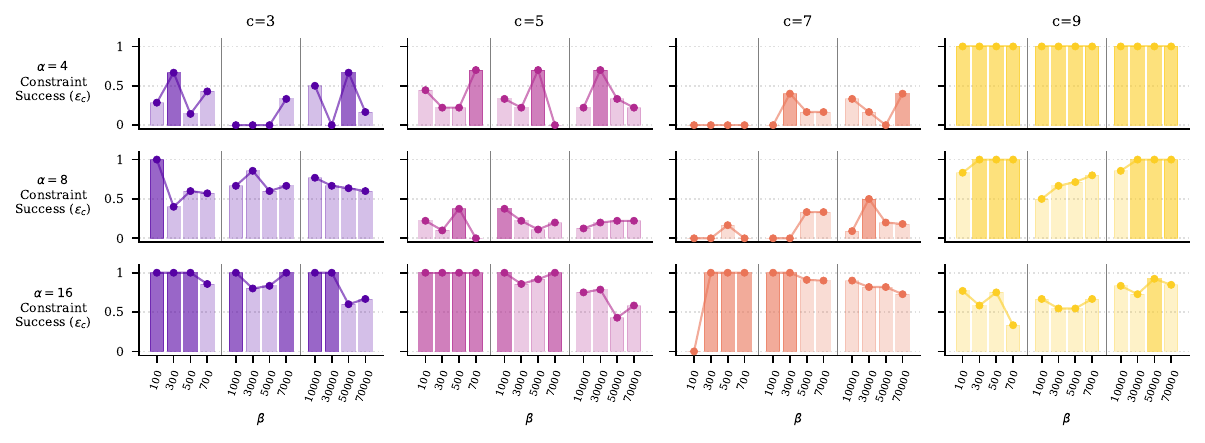}
                    \caption{Constraint succeess ratio $\varepsilon_c$ for different configurations of ($\alpha$, $\beta$, $c$) on an 18-node graph. From top to bottom, each row corresponds to $\alpha \in \{4, 8, 16\}$, while each column corresponds to $c \in \{3, 5, 7, 9\}$. At fixed $\alpha$ and $c$, $\varepsilon_c$ is reported for three ranges of $\beta$: $\beta \sim 10^2$, $\beta \sim 10^3$, and $\beta \sim 10^4$. The highlighted bars indicate the highest $\varepsilon_c$ achieved for each $(\alpha, c)$ configuration. Each bar corresponds to the mean ratio over 10 simulations.}
                    \label{fig: beta performance}
                \end{figure*}
                The parameter $\beta$ controls the strength of the penalty applied when the cut violates the constraint, i.e., when the solution does not separate exactly $c$ nodes. Selecting appropriate values for penalty parameters is a well-known challenge in combinatorial optimization and remains an open problem in both classical and quantum optimization methods \cite{diez_garcia_exact_2022}.
                We have tested the performance of the PCE for several combinations of the parameters $\alpha$,$\beta$ and the constraint parameter $c$. Figure~\ref{fig: beta performance} illustrates the results. The main observations can be summarized as follows:
                \begin{enumerate}
                    \item For each value of $c$, the overall performance of PCE is primarily governed by the choice of $\alpha$.
                    \item For fixed values of $\alpha$ and $c$, there is no clear range of $\beta$ values that consistently yields optimal performance. In practice, similar performance levels are observed for $\beta \sim 10^2$, $\beta \sim 10^3$, and $\beta \sim 10^4$.
                    \item At fixed $\alpha$ and $c$, the performance is highly sensitive to the specific value of $\beta$ within the same order of magnitude. For example, for $\alpha = 4$ and $c=5$, a peak in the constraint success ratio $\varepsilon_c$ is observed at $\beta = 5000$, whereas a slight increase to $\beta = 7000$ causes $\varepsilon_c$ to drop to zero. Several analogous cases are visible in the figure.
                    \item The value of $\beta$ that yields the best performance is not transferable across different $(\alpha, c)$ configurations. This is evident from the highlighted bars in Figure~\ref{fig: beta performance}, which indicate the highest $\varepsilon_c$ for each configuration. For instance, while $\beta = 3000$ and $\beta = 70000$ provide the best results for $\alpha = 4$ and $c = 7$, these same values lead to significantly poorer performance for $c = 3$.
                \end{enumerate}

                Therefore, while $\alpha$ determines the overall performance regime of PCE, the optimal performance for a given configuration $(\alpha, c)$ is critically dependent on the choice of $\beta$. If $\beta$ is not appropriately tuned, the algorithm fails to satisfy the constraint and produces suboptimal solutions. Moreover, since the optimal value of $\beta$ varies with $c$, these observations motivate the search for a functional dependence $\beta(c)$.
                
\begin{figure*}[t]
                    \includegraphics[width=\linewidth]{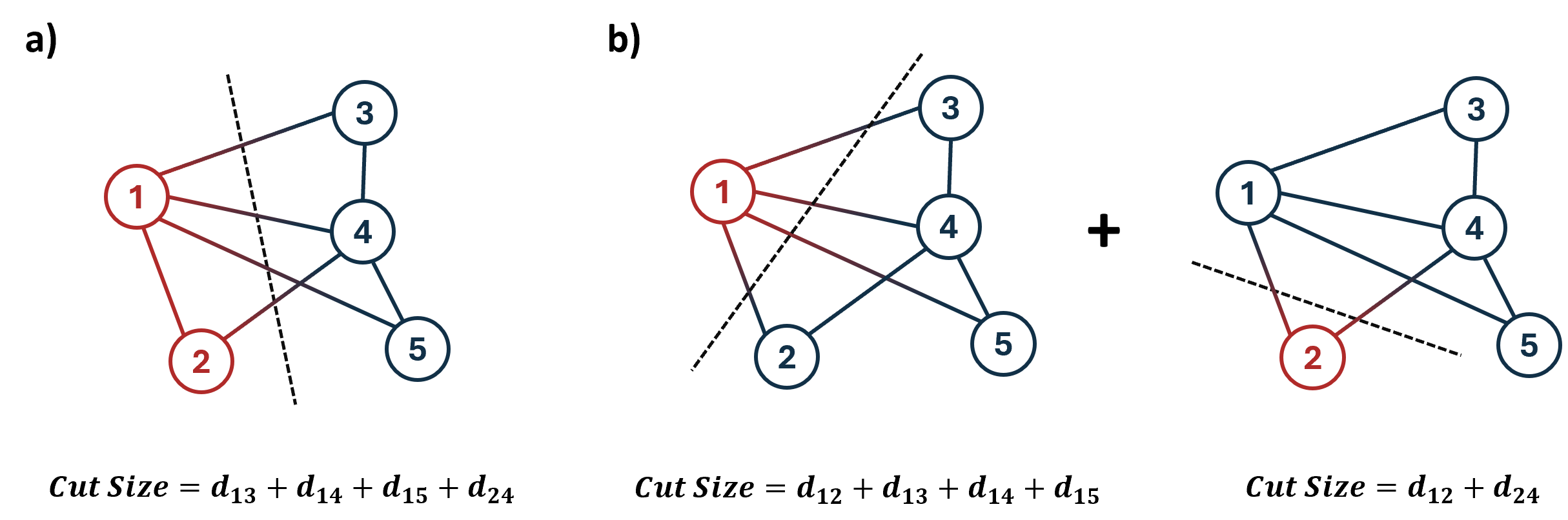}
                    \caption{Example of a cut separating two nodes ($c=2$). The cut size of the configuration (a) is less than or equal to the sum of the cut sizes obtained by separating each node individually (b).}
                    \label{fig: cut example c=2}
                \end{figure*}

                We propose a heuristic method to estimate an effective penalty parameter $\beta(c) = \beta_c$. In general, $\beta$ must be chosen in accordance with the scale of the loss function to properly balance the constraint term against the cut objective. Before introducing the heuristic used to compute $\beta_c$, we consider a simple illustrative example: a cut with $c = 2$ in a five-node graph, shown in Figure~\ref{fig: cut example c=2}. 
                \begin{figure*}[t]
                    \includegraphics[]{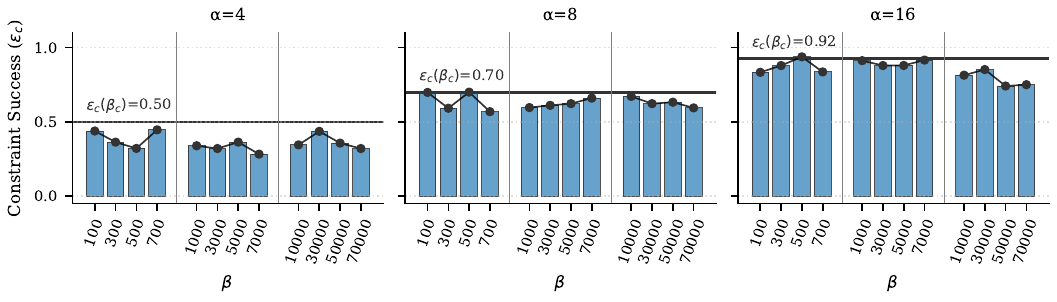}
                    \caption{Constraint success ratio $\varepsilon_c$ for different configurations of $\{\alpha,\beta\}$ and $\{\alpha,\beta_c\}$ on the 18-node graph. Each bar represents the mean value of $\varepsilon_c$ averaged over all executions and $c$ values.}
                    \label{fig: beta_c performance}
                \end{figure*}
                From Figure~\ref{fig: cut example c=2}, the cut size obtained by separating nodes ${1,2}$ from nodes ${3,4,5}$ is strictly smaller than the sum of the cut sizes obtained by separating nodes $1$ and $2$ individually. This relationship can be expressed by the following inequality:
                \begin{equation} 
                d_{13}+d_{14}+d_{23}+d_{24} \le d_{13}+d_{14}+d_{23}+d_{24}+2\,d_{12} 
                \label{eq: example_cut_inequality} 
                \end{equation}

                It is important to note that this expression changes when $c = 3$, since the set of negative variables becomes $z_3 = z_4 = z_5 = -1$, even though the resulting cut may remain unchanged. For this reason, we restrict the formulation to the regime $c \in [2, m/2]$. 
                Let us introduce the quantity
                \begin{equation}
                d_i = \sum_{j=1}^{n} d_{ij},
                \end{equation}
                which represents the sum of all weights connected to node $i$. For a fixed value of $c$ in the constraint, let ${\bar{d}_i}$ $|$ ${i \in n_c}$ denote the set of the $c$ largest values of $d_i$, where $n_c$ indexes the corresponding nodes. Then, the following expression is always true:
                \begin{equation}
                    \sum_{i}^{n-1}\sum_{j>i}^{n}{\frac{1}{2}d_{ij}\left(1-z_iz_j\right)}\ \le\ \ \sum_{i}^{n}{\frac{1}{2}d_i\left(1-z_i\right)}\le\sum_{i}^{n_c}{\bar{d}}_i
                \end{equation}
                The full mathematical deduction can be found on Appendix. This upper bound holds for any value of $c$ and for any graph. Based on this result, we propose to define the penalty parameter as
                \begin{equation}
                \beta(c) = \beta_c = \sum_{i \in n_c} \bar{d}_i.
                \end{equation}
                We have tested this proposed value of $\beta_c$. In Figure~\ref{fig: beta_c performance} it is shown a comparison between the constraint success ratio $\varepsilon_c$ obtained by different configurations of $\alpha$, $\beta$, and the proposed value $\alpha$, $\beta_c$. From the results, two main conclusions can be drawn:
                \begin{enumerate}
                \item The overall performance is predominantly determined by $\alpha$. This trend is observed across the full range of $\beta$ values and is also reflected in the results obtained using $\beta_c$, where $\varepsilon_c$ increases from $0.50$ at $\alpha = 4$ to $0.92$ at $\alpha = 16$.
                \item The proposed penalty parameter $\beta_c$ yields equal or improved performance using a single value for all $c$, independently of $\alpha$.
                \end{enumerate}
                In conclusion, the proposed expression for $\beta(c)$ yields consistent performance and can therefore be used to study the algorithm in a systematic manner, without being affected by performance drops associated with particular choices of $\beta$ for specific $(\alpha, c)$ configurations.  Moreover, these results indicate that $\alpha$ is the dominant parameter governing the overall performance of the algorithm.

            \subsection{Regularization term}
                                \begin{figure*}[t]
                    \includegraphics[]{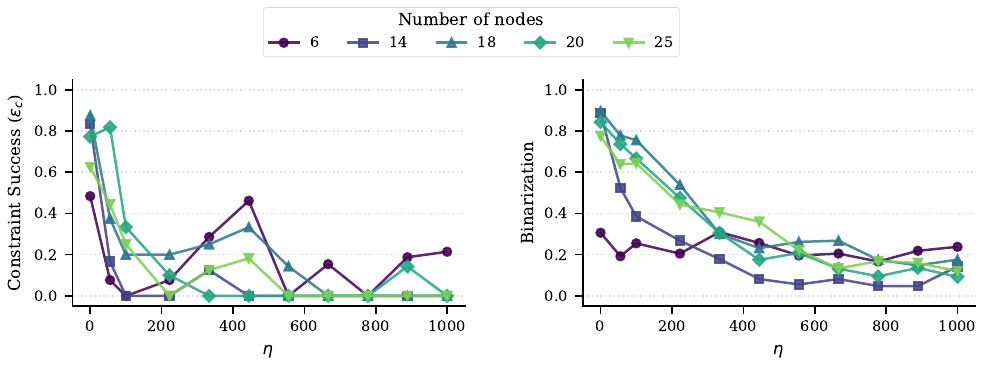}
                    \caption{Effect of the regularization strength $\eta$ on constraint satisfaction and binarization. Each data point corresponds to the mean value averaged over all $c \in [2, n/2]$, 5 times executed each one.}
                    \label{fig: regularization}
                \end{figure*}
                The regularization term is introduced in the original paper as a quadratic penalty centered at zero. The term is designed to keep the expectation values far from large values, which is reported to promote smoother optimization dynamics by effectively explore a broader range of variable configurations. Its formulation is:
                \begin{equation}
                    L^{reg}=\eta\left[\frac{1}{n}\sum_{i}^{n}{\tanh\left(\alpha\left\langle\Pi_i\right\rangle\right)^2}\right]^2
                    \label{eq: regularization term}
                \end{equation}
                where $n$ denotes the number of nodes and acts as a normalization factor, and $\eta > 0$ controls the strength of the regularization. In the original PCE paper, $\eta$ is analytically estimated for the Max-Cut problem; however, in more general settings it is typically chosen to be of the same order of magnitude as the loss function. 
                In our problem, this term may interfere with constraint satisfaction, as the latter requires fully binarized variables. To assess its impact, we have performed simulations across a range of values of $\eta$ for several graph instances. As it is shown in the results on Figure~\ref{fig: regularization}, as the strenth $\eta$ used in the regularization term, the constraint success ratio decreases. In parallel, the binarization of the variables also decreases, as expected.

                Therefore, we conclude that the regularization term can be counterproductive for constrained problems using the PCE.

    \section{Progressive Binarization PCE}
    \label{sec_iterative_PCE}
        
        As discussed in the previous sections, the binarization parameter $\alpha$ must be sufficiently large to induce binarization of the encoded variables, yet not so large that the optimization stalls in the plateau regime of the $\tanh$ function. Rather than fixing $\alpha$ \emph{a priori}, here we determine it dynamically: 
        starting from a small initial value, $\alpha$ is progressively increased so that the encoded variables are gradually driven towards the binary domain as the optimization proceeds. We refer to this method as \emph{Progressive Binarization-PCE} (PB-PCE).

        \subsection{Continuation formulation}
        \label{subsec_continuation}

            PB-PCE can be cast as an adaptive continuation (graduated-optimization) method~\citep{blake_visual_1987, allgower_numerical_1990, mobahi_theoretical_2015, hazan_graduated_2016}. The binary problem underlying PCE can be written as
            \begin{equation}
                L(\vec{\theta}) = f\!\left(\mathrm{sign}(\langle\Pi_1\rangle),\ldots,\mathrm{sign}(\langle\Pi_n\rangle)\right),
            \end{equation}
            which is hard to optimize because the sign function is discontinuous. PCE replaces the sign by a smooth relaxation, defining the one-parameter family of problems
            \begin{equation}
                L_\alpha(\vec{\theta}) = f\!\left(\tanh(\alpha\langle\Pi_1\rangle),\ldots,\tanh(\alpha\langle\Pi_n\rangle)\right),
                \label{eq: smoothed family}
            \end{equation}
            where $\alpha$ acts as the continuation parameter controlling the binarization of the encoded variables. Since
            \begin{equation}
                \lim_{\alpha\to\infty}\tanh(\alpha x) = \mathrm{sign}(x), \qquad x\neq 0,
            \end{equation}
            the relaxed loss converges pointwise to the objective, $L_\alpha(\vec{\theta})\to L(\vec{\theta})$ as $\alpha\to\infty$ for $\langle\Pi_i\rangle\neq 0$. At a small initial value $\alpha_0$ the landscape is smooth and easy to optimize, whereas in the limit $\alpha\to\infty$ the original binary constrained problem is recovered. PB-PCE follows a minimizer $\vec{\theta}^\star(\alpha)$ along a sequence of increasing values $\alpha_0<\alpha_1<\cdots$, re-optimizing the circuit parameters from the previously converged configuration at each stage, thereby tracking the solution from the smooth relaxation towards the binary domain.

            Unlike a continuation method with a prefixed schedule $\alpha_1<\alpha_2<\cdots<\alpha_N$, the trajectory in PB-PCE is \emph{adaptive}: the next value of the continuation parameter depends on the current state of the encoded variables,
            \begin{equation}
                \alpha_{k+1} = g\!\left(\alpha_k,\{\langle\Pi_i\rangle\}\right),
                \label{eq: adaptive trajectory}
            \end{equation}
            as specified by the update rule in Algorithm~1. This state-dependent schedule acts as an adaptive step-size control: the increment of $\alpha$ is chosen so that the least-binarized variable is pushed just past the binarization threshold, which prevents both abrupt jumps into the plateau regime and stalling of the optimization. Conceptually, the same mechanism underlies the progressive binarization of continuous relaxations in the training of binary neural networks, where a smooth surrogate of the sign function is gradually sharpened until it coincides with the sign function~\citep{courbariaux_binaryconnect_2015, lahoud_self-binarizing_2019}.

        \subsection{Algorithm}
        \label{subsec_algorithm}

            The core idea is to let the PCE optimization converge for an initial value of $\alpha$, and then, based on the result, compute the next value of $\alpha$. The optimization is subsequently restarted from the final parameter configuration of the parametrized quantum circuit using this larger value of $\alpha$. By repeating this process the variables are progressively pushed toward full binarization without abruptly entering the plateau regime of the loss landscape. In detail, the proposed heuristic for adjusting $\alpha$ operates as described in Algorithm 1.
        
        \begin{figure}[!t]
        \refstepcounter{myalgorithm}
        \hrule
        \vspace{0.2em}
        \textbf{Algorithm \themyalgorithm: PB-PCE heuristic}
        \vspace{0.2em}
        \hrule
        \vspace{0.2em}
        \begin{algorithmic}[1]
            \label{alg:adaptive_alpha}
            \Require Initial value $\alpha_0$, threshold $M$
            \State $\alpha \gets \alpha_0$
            \Repeat
                \State Run PCE until convergence with current $\alpha$
                \State Identify the set $\mathcal{I} = \{ i \mid |\tanh{(\alpha\langle\Pi_i\rangle)}| < M \}$
                \If{$\mathcal{I} \neq \emptyset$}
                    \State Select
                    \[
                    i^\star = \arg\min_{i \in \mathcal{I}}
                    \left| \left|\tanh{(\alpha\langle\Pi_i\rangle)}\right| - M \right|
                    \]
                    \State Update
                    \[
                    \alpha \gets \alpha
                    \frac{\operatorname{arctanh}(M)}
                         {\operatorname{arctanh}(|\tanh({\alpha \langle\Pi_{i^\star}\rangle})|)}
                    \]
                \EndIf
            \Until{$|\tanh{(\alpha\langle\Pi_i\rangle)}| \ge M \;\; \forall i$}
            \vspace{0.2em}
            \hrule
            \end{algorithmic}
        \end{figure}

        \setlength{\tabcolsep}{7pt}  
        \begin{table}[t]
            \centering
            \small
                \begin{tabular}{r c c r}
                \toprule
                Nodes &
                \multicolumn{2}{c}{Constraint satisfied} &
                Runs (\%)\\
                \cmidrule(lr){2-3}
                 & PB-PCE & PCE($\alpha_f$) & \\
                \midrule
                \multirow{4}{*}{6}
                  & \checkmark & \checkmark & 53 \% \\
                  & \checkmark & $\times$ & 47 \% \\
                  &$\times$& \checkmark & 0 \% \\
                  &$\times$&$\times$& 0 \% \\
                \midrule
                \multirow{4}{*}{14}
                  & \checkmark & \checkmark & 45 \% \\
                  & \checkmark &$\times$& 55 \% \\
                  &$\times$& \checkmark & 0 \% \\
                  &$\times$&$\times$& 0 \% \\
                \midrule
                \multirow{4}{*}{18}
                  & \checkmark & \checkmark & 37 \% \\
                  & \checkmark &$\times$& 63 \% \\
                  &$\times$& \checkmark & 0 \% \\
                  &$\times$&$\times$& 0 \% \\
                \midrule
                \multirow{4}{*}{20}
                  & \checkmark & \checkmark & 29 \% \\
                  & \checkmark &$\times$& 71 \% \\
                  &$\times$& \checkmark & 0 \% \\
                  &$\times$&$\times$& 0 \% \\
                \midrule
                \multirow{4}{*}{25}
                  & \checkmark & \checkmark & 22 \% \\
                  & \checkmark &$\times$& 78 \% \\
                  &$\times$& \checkmark & 0 \% \\
                  &$\times$&$\times$& 0 \% \\
                \bottomrule
                \end{tabular}
            \caption{Constraint satisfaction outcomes for PB-PCE and PCE($\alpha_f$).
            Each row reports the percentage of executions corresponding to each combination of outcomes.
            $\checkmark$ indicates that the constraint is satisfied, while $\times$ denotes violation.}
            \label{tab: contingency_constraint}
            \end{table}    
        \begin{table}[t]
            \centering
            \small
                \begin{tabular}{r c c r}
                \toprule
                Nodes &
                \multicolumn{3}{c}{CutSize} \\
                \cmidrule(lr){2-4}
                 & PB-PCE & PCE & $\Delta$ (\%) \\
                \midrule
                6 & 1.0620 & 1.1589 & -8.36 \% \\
                \addlinespace
                14 & 1.3195 & 1.8265 & -27.76 \% \\
                \addlinespace
                18 & 1.6523 & 2.5674 & -35.65 \% \\
                \addlinespace
                20 & 1.1526 & 1.2543 & -8.11 \% \\
                \addlinespace
                25 & 1.0999 & 1.1666 & -5.72 \% \\
                \bottomrule
                \end{tabular}
            \caption{Mean CutSize achieved (lower is better), computed over matched pairs where both methods satisfy the constraint.
            $\Delta$: percentage difference of PB-PCE relative to PCE.}
            \label{tab: cut_size_comparison}
        \end{table}
        \begin{figure*}[t]
            \centering
            \includegraphics[width=\textwidth]{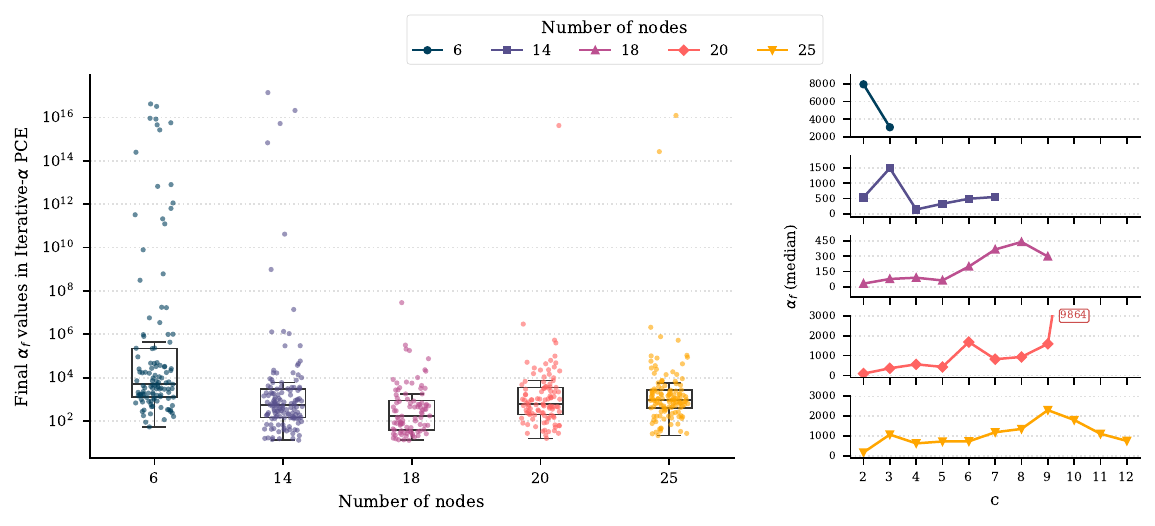}
            \caption{Final values of $\alpha$ reached by PB-PCE.
                    Left: distribution of $\alpha$ across executions for different graph sizes.
                    Right: median $\alpha$ as a function of the constraint parameter $c$, with interquartile ranges.}
            \label{fig: distribution of alphas}
        \end{figure*}
        
        We have tested PB-PCE on graphs with $n=\{6,14,18,20,25\}$ nodes. Details of both the initial value of $\alpha$ and the threshold value $M$ can be found on Appendix. To disentangle the effect of the final value of $\alpha$ from that of the iterative procedure itself, we perform an additional control experiment. For each execution of PB-PCE, we also run a standard PCE using a fixed value $\alpha_f$, where $\alpha_f$ corresponds to the final value reached by the iterative scheme in that same execution. This paired comparison allows us to assess whether any observed performance improvement arises solely from reaching a suitable value of $\alpha$, or whether the iterative update mechanism provides an intrinsic advantage. Table~\ref{tab: contingency_constraint} reports, for each graph size, the percentage of runs falling into each possible combination of constraint satisfaction outcomes for the two methods. 
        
        If both approaches were effectively equivalent, the fraction of runs in which only one of them satisfies the constraint would be close to zero. The results clearly deviate from this scenario. In all tested graph sizes, the percentage of runs in which the PCE satisfies the constraint while PB-PCE does not is $0\%$. Conversely, when PB-PCE satisfies the constraint, PCE does it only in a fraction of the runs, ranging from $53\%$ for $n=6$ down to $22\%$ for $n=25$. Notably, the proportion of executions in which PB-PCE succeeds while the PCE fails increases systematically with the graph size, reaching $78\%$ for $n=25$.

        Regarding the CutSize, in Table~\ref{tab: cut_size_comparison} are shown the results for only those simulations in which both methods satisfy the constraint. PB-PCE consistently achieves a better solution quality, yielding cut sizes between a $5\%-30\%$ smaller than the PCE. 

        Taken together, these results indicate that the observed performance gains cannot be attributed solely to the final value $\alpha_f$ achieved by the algorithm, but rather to the iterative evolution of $\alpha$ during the optimization process itself.
        \begin{figure*}[t]
            \centering
            \includegraphics[width=\textwidth]{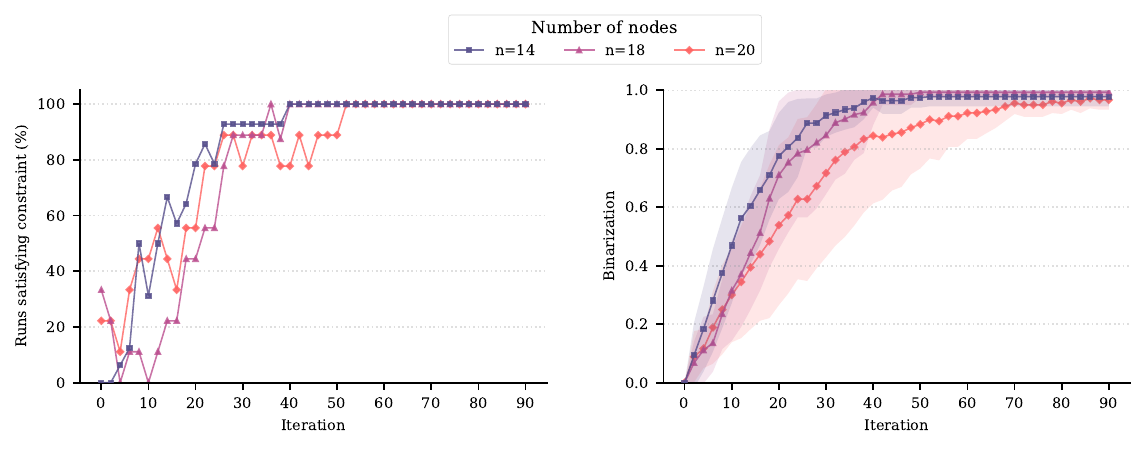}
            \caption{Evolution of constraint satisfaction and binarization during PB-PCE as a function of the iteration, i.e. of each update of $\alpha$. Left: percentage of runs that satisfy the constraint. Right: binarization metric, where the shaded regions indicate the dispersion across runs. Both quantities saturate at essentially the same iteration.}
            \label{fig: corte y binarizacion vs iteration}
        \end{figure*}

        Figure~\ref{fig: distribution of alphas} shows the distribution of final values $\alpha_f$ reached across all executions of PB-PCE. No clear concentration around a specific value of $\alpha_f$ is observed. Instead, the distribution spans several orders of magnitude, ranging approximately from $10^{2}$ to $10^{4}$, with a subset of executions reaching values as large as $10^{6}$ and even up to $10^{16}$. Consistently, when inspecting the median value of $\alpha_f$ as a function of the constraint strength $c$ and graph size, no systematic trend emerges. This behavior indicates that PB-PCE does not converge toward a specific optimal value of $\alpha$, but rather explores a broad range of final parameters depending on the instance and execution.
        Evaluating the performance of the PCE over this ensemble of final values provides a representative assessment of the behavior of standard PCE across a wide and heterogeneous range of $\alpha$. Table~\ref{tab: iterative_vs_final_alpha} summarizes the aggregated results obtained from all executions, reporting constraint satisfaction ratio $\varepsilon_c$, binarization, and cut size for both PB-PCE and the corresponding PCE($\alpha_f$).
        \setlength{\tabcolsep}{11pt}  
        \begin{table}[hpbt]
            \centering
            \small
                \begin{tabular}{r l r r r}
                    \toprule
                    Nodes & PCE Type & $\varepsilon_{c}$ & Bin. & CutSize \\
                    \midrule
                    6  & PB-PCE        & 1.00 & 1.00 & 1.05 \\
                    6  & PCE               & 0.53 & 0.99 & 1.16 \\
                    \addlinespace
                    14 & PB-PCE        & 1.00 & 1.00 & 1.41 \\
                    14 & PCE               & 0.45 & 0.93 & 1.93 \\
                    \addlinespace
                    18 & PB-PCE        & 1.00 & 1.00 & 1.80 \\
                    18 & PCE               & 0.37 & 0.93 & 2.57 \\
                    \addlinespace
                    20 & PB-PCE        & 1.00 & 1.00 & 1.12 \\
                    20 & PCE               & 0.29 & 0.51 & 1.25 \\
                    \addlinespace
                    25 & PB-PCE        & 1.00 & 1.00 & 1.13 \\
                    25 & PCE               & 0.22 & 0.96 & 1.17 \\
                    \bottomrule
                \end{tabular}
            \caption{Results using PB-PCE and the PCE with the final $\alpha_f$ values obtained from each run.}
            \label{tab: iterative_alpha_results}
        \end{table}
        
        The results are unambiguous. PB-PCE achieves $100\%$ constraint satisfaction and full binarization across all tested graph sizes. In contrast, while the PCE consistently reaches binarized solutions, constraint satisfaction is achieved only in a fraction of the executions, decreasing from $53\%$ for small graphs down to $22\%$ for the largest instances. This degradation with graph size is consistent with the stagnation effects observed in the standard PCE analysis. Moreover, PB-PCE also yields superior solution quality, achieving smaller cut sizes overall.

        Intuitively, this behavior can be understood as follows. As $\alpha$ increases, a subset of variables enters the plateau region of the $\tanh$ function, while the remaining variables stay within its linear regime. As discussed previously, variables in the linear regime allow the optimizer to reduce the loss by minimizing the constraint term without truly satisfying it, while also reducing the cut size in the continuous domain. When $\alpha$ is increased and the optimization is restarted from the previously converged solution, the width of the plateau region expands (Figure~\ref{fig: tanh and derivate}). Consequently, variables that were already binarized become more deeply embedded in the plateau, making it increasingly difficult for the optimizer to move them back into the linear regime. As a result, the optimization dynamics naturally focus on the remaining variables in the linear regime that can still explore the loss landscape. As $\alpha$ continues to grow, more variables transition into the plateau region, and fewer variables remain able to explore continuous configurations. This progressively reduces the  values available to minimize the constraint without actually satisfying it. In this way, the effective continuous search space explored by the optimizer shrinks gradually, biasing the optimization toward configurations that fulfill the constraint. This mechanism explains why increasing $\alpha$ progressively leads to a full constraint success ratio and full binarization of the variables. In the continuation picture of Section~\ref{subsec_continuation}, this is precisely why neither endpoint of the family in Eq.~(\ref{eq: smoothed family}) can be solved directly: at small $\alpha$ the relaxation lies too far from the binary problem and the optimizer settles on fractional, infeasible configurations, whereas a large-$\alpha$ instance solved from a cold start stalls in the $\tanh$ plateaus. Tracking the minimizer along increasing $\alpha$ is what threads between these two failure modes. 
        
        Figure~\ref{fig: corte y binarizacion vs iteration} illustrates this process by showing, as a function of the iteration (i.e. of each update of $\alpha$), the percentage of runs that satisfy the constraint together with the binarization metric. Both quantities follow remarkably similar trajectories: for each graph, the iteration at which all runs satisfy the constraint essentially coincides with the one at which the variables reach full binarization. The agreement also holds across instances, since the graph that binarizes more slowly is likewise the last one to attain complete constraint satisfaction. This provides direct evidence that, in PB-PCE, constraint satisfaction is governed by the \textbf{progressive binarization} of the encoded variables. An example of the evolution of the variables during the iterative process is provided in the Appendix. 
                
        In conclusion, the proposed PB-PCE method provides an optimization subroutine that does not require any prior estimation of the parameter $\alpha$. By progressively updating $\alpha$ throughout the optimization process, it consistently achieves high-quality solutions, driving the optimization toward the binary domain through full variable binarization while satisfying the constraint in $100\%$ of the executions and systematically outperforming standard PCE.
        
        From a hardware perspective, this iterative strategy increases the number of circuit executions required. However, the quantum circuit remains unchanged across iterations, and only the parameters of the variational ansatz are updated.

    \subsection{Large-scale simulation}

        Due to the outstanding results using PB-PCE, we scaled up the size of the target graphs to $n\in\{50, 150, 300\}$. Table~\ref{fig_large_scale_graph_info} summarizes the structural properties of all graph instances considered in this section. For these larger instances, we move from the Qiskit-based implementation to executing the complete PB-PCE pipeline using the \textit{Fujitsu QARP} framework, which provides an integrated and highly optimized implementation of the method. For reference, Appendix~\ref{appendix_runtime} presents a runtime comparison between the QARP framework and our Qiskit-based implementation. While we do not claim that the Qiskit implementation is fully optimized for this problem, QARP is an in-house framework that enables deeper optimization and tighter workflow integration, resulting in significantly improved execution times.
    
        \setlength{\tabcolsep}{4pt}  
        \begin{table}[hbpt]
        \centering
        \begin{tabular}{lrrrc}
        \toprule
        Nodes & Edges & Strength & Strength Std & Connected\\
        \midrule
        25 & 300 & 22.00 & 5.91 & \checkmark \\
        50  & 1223 & 27.75  & 9.33 & \checkmark \\
        150 & 11175 & 49.41 & 11.55 & \checkmark \\
        300 & 44850 & 50.00 & 20.41 & \checkmark \\
        \bottomrule
        \end{tabular}
        \caption{Key structural properties of the analyzed graphs. Mean Strength denotes the average node strength, computed as the mean incident edge weight per node. Strength Std denotes the average standard deviation of incident edge weights across nodes. The final column indicates whether each graph forms a single connected component.}
        \label{fig_large_scale_graph_info}
        \end{table}

        In this section the heuristic to change the $\alpha$ parameter was slightly changed from Algorithm 1 (line 7). Due to the larger number of nodes in these graphs, we frequently observed variables reaching values too close to the threshold $M$ of the heuristic. This resulted in increments of $\alpha$ that were too small and stalled the optimization. To ensure $\alpha$ is increased we changed the update of $\alpha$ between iterations to:
        \begin{equation}
            \alpha \gets \alpha
                    \frac{\operatorname{arctanh}(M)}
                         {|\tanh({\alpha \langle\Pi_{i^\star}\rangle})|}
        \end{equation}
        A detailed comparison of both heuristics is provided in Appendix~\ref{appendix_heuristics}.


    \subsection{PB-PCE large-scale results}

        \setlength{\tabcolsep}{7pt}  
        \begin{table}[hbpt]
        \centering  
        \small
        \begin{tabular}{r c c r}
        \toprule
        Nodes &
        \multicolumn{2}{c}{Constraint satisfied} &
        Runs (\%) \\
        \cmidrule(lr){2-3}
         & PB-PCE & PCE($\alpha_f$) & \\
        \midrule
        \multirow{4}{*}{25}
         & \checkmark & \checkmark & 16 \% \\
         & \checkmark & $\times$  & 84 \% \\
         & $\times$  & \checkmark & 0 \% \\
         & $\times$  & $\times$  & 0 \% \\
        \midrule
        \multirow{4}{*}{50}
         & \checkmark & \checkmark & 20 \% \\
         & \checkmark & $\times$  & 80 \% \\
         & $\times$  & \checkmark & 0 \% \\
         & $\times$  & $\times$  & 0 \% \\
        \midrule
        \multirow{4}{*}{150}
         & \checkmark & \checkmark & 8 \% \\
         & \checkmark & $\times$  & 87 \% \\
         & $\times$  & \checkmark & 0 \% \\
         & $\times$  & $\times$  & 5 \% \\
        \midrule
        \multirow{4}{*}{300}
         & \checkmark & \checkmark & 25 \% \\
         & \checkmark & $\times$  & 63 \% \\
         & $\times$  & \checkmark & 0 \% \\
         & $\times$  & $\times$  & 12 \% \\
        \bottomrule
        \end{tabular}
        \caption{Constraint satisfaction outcomes for PB-PCE and PCE($\alpha_f$).
            Each row reports the percentage of executions corresponding to each combination of outcomes.
            $\checkmark$ indicates that the constraint is satisfied, while $\times$ denotes violation.}
        \label{tab: iterative_vs_final_alpha}
        \end{table}

        \begin{table}[hbpt]
        \centering
        \small
        \begin{tabular}{r r r r c}
        \toprule
        Nodes &
        \multicolumn{3}{c}{CutSize} & PB-PCE \\
        \cmidrule(lr){2-4}
         & PB-PCE & PCE & $\Delta$ (\%) & Iterations \\
        \midrule
        25 & 1.0290 & 1.1690 & -11.97\% & 19 \\
        \addlinespace
        50 & 1.0932 & 1.2407 & -11.89\% & 16 \\
        \addlinespace
        150 & 1.0722 & 1.1075 & -3.19\% & 16 \\
        \addlinespace
        300 & 1.0001 & 1.0077 & -0.75\% & 17 \\
        \bottomrule
        \end{tabular}
        \caption{Mean CutSize achieved (lower is better), computed over matched pairs where both methods satisfy the constraint.
            $\Delta$: percentage difference of PB-PCE relative to PCE. The last column reports the average number of PB-PCE iterations, i.e. of updates of $\alpha$.}
        \label{tab: cut_size_comparison_1}
        \end{table}

        \setlength{\tabcolsep}{4pt}  
        \begin{table}[!t]
        \centering
        \small
        \begin{tabular}{r r r r c}
        \toprule
        Nodes & Qubits &
        \multicolumn{2}{c}{Constraint Success} &
        $\Delta$ CutSize \\
        \cmidrule(lr){3-4}
         & & PB-PCE & PCE & PB-PCE vs PCE \\
        \midrule
        6   & 3 & 100 $\%$ & 53 $\%$ & -8.36 $\%$ \\
        \addlinespace
        14  & 4 & 100 $\%$ & 45 $\%$ & -27.76 $\%$ \\
        \addlinespace
        18  & 4 & 100 $\%$ & 37 $\%$ & -35.65 $\%$ \\
        \addlinespace
        20  & 5 & 100 $\%$ & 29 $\%$ & -8.11 $\%$ \\
        \addlinespace
        25  & 5 & 100 $\%$ & 16 $\%$  & -11.97 $\%$ \\
        \addlinespace
        50  & 6 & 100 $\%$  & 20 $\%$ & -11.89 $\%$ \\
        \addlinespace
        150 & 8 & 94 $\%$ & 13 $\%$ & -3.19 $\%$ \\
        \addlinespace
        300 & 9 & 88 $\%$ & 25 $\%$ & -0.75 $\%$ \\
        \bottomrule
        \end{tabular}
        \caption{Summary of PB-PCE results. $\Delta$ CutSize denotes the percentage difference in cut size relative to the standard PCE. Negative values indicate that PB-PCE achieves a smaller cut size than standard PCE.}
        \label{tab:iterative_summary}
        \end{table}

        The same analysis as for the smaller graph instances was performed. Constraint satisfaction results are reported in Table~\ref{tab: iterative_vs_final_alpha} and cut sizes are presented in Table~\ref{tab: cut_size_comparison_1}. 
        The results are clear, PB-PCE consistently outperforms the standard PCE, with almost 100$\%$ of constraint satisfaction, while also yielding better cut sizes. The results for several values of the constraint parameter $c$ are reported in the Appendix.

        Furthermore, the last column of Table~\ref{tab: cut_size_comparison_1} reports the average number of iterations (defined as updates of the binarization parameter $\alpha$) for each graph instance. Interestingly, the number of iterations seems to be independent of the graph size.

        To conclude, we present in Table~\ref{tab:iterative_summary} a summary of the results obtained in all the graphs studied in this paper.


\section{Conclusions and future work}
\label{sec_conclusions}

    In this work, we conducted an in-depth study of the PCE approach for a constrained combinatorial optimization problem. Our analysis identified that the binarization of the variables plays a crucial role in obtaining valid solutions that satisfy the constraint term. Without careful optimization of the hyperparameters, the standard PCE struggles to enforce constraints, which often leads to infeasible solutions.

    To address this limitation, we introduced PB-PCE, a modification of the baseline designed to handle constrained optimization problems more effectively. This approach achieves near-100 $\%$ constraint satisfaction while consistently producing better cut sizes compared to the standard PCE. Notably, using this strategy, we were able to solve constrained graph instances with up to 300 nodes using only 9-qubits quantum circuits. 
    We believe this represents a significant step forward in the application of PCE to large-scale constrained optimization problems, enabling the algorithm to be applied to more realistic problem instances while requiring a reduced number of qubits. This characteristic makes the approach particularly well suited to the NISQ era.

    Nevertheless, this improvement comes at the cost of increased computational overhead. The iterative scheme typically requires on the order of 10–20 executions of individual PCE runs due to repeated updates of the parameter $\alpha$. Importantly, this overhead appears to be largely independent of the number of nodes, which makes the algorithm promising from a scalability perspective. Future research should therefore focus on developing more effective heuristics to better understand the scalability of the method, as well as on reducing the number of required iterations while still ensuring constraint satisfaction. Another promising direction is the exploration of advanced encoding strategies, such as single-Pauli correlation schemes, which could reduce the total number of required quantum circuits by up to a factor of three. Together, these findings may further enhance the practicality of the PCE algorithm for constrained optimization on near-term quantum hardware.

\bibliography{references}
\clearpage
\appendix
\label{appendix}
\onecolumngrid

\section{$\beta_c$ calculous}

We introduce here the full mathematical development of the $\beta(c)$ expresion. Let $x_i\in\{-1,1\}$. For convenience, we first rewrite the CutSize term as:
               \begin{equation}
                \begin{split}
                \sum_{i=1}^{n-1}\sum_{j>i}^{n}\frac{1}{2}d_{ij}\left(1-x_ix_j\right)
                &= \sum_{i=1}^{n-1}\sum_{j>i}^{n}\frac{1}{2}d_{ij}\left(1-x_ix_j+x_i-x_i\right) = \\
                &= \sum_{i=1}^{n-1}\sum_{j>i}^{n}\frac{1}{2}d_{ij}\left(1-x_i\right)
                   + \sum_{i=1}^{n-1}\sum_{j>i}^{n}\frac{1}{2}d_{ij}x_i\left(1-x_j\right)
                \end{split}
                \end{equation}
                given $d_{ij}=d_{ji}$, then:
                \begin{equation}
                    \sum_{i}^{n-1}\sum_{j>i}^{n}{\frac{1}{2}d_{ij}\left(1-x_ix_j\right)}\ = \\ =\sum_{i}^{n-1}\sum_{j\neq\ i}^{n}{\frac{1}{4}d_{ij}\left(1-x_i\right)}+\sum_{i}^{n-1}\sum_{j\neq\ i}^{n}{\frac{1}{4}d_{ij}x_i\ \left(1-x_j\right)}
                \end{equation}
                now, the second term:
                \begin{equation}
                        \sum_{i}^{n-1}\sum_{j\neq\ i}^{n}{\frac{1}{4}d_{ij}\left(1-x_i\right)\left(1-x_j\right)}
                        =\sum_{i}^{n-1}\sum_{j\neq\ i}^{n}{\frac{1}{4}d_{ij}\left(1-x_i-x_j+x_ix_j\right)}=\sum_{i}^{n-1}\sum_{j\neq\ i}^{n}{\frac{1}{4}d_{ij}\left(1-x_i\right)}-\sum_{i}^{n-1}\sum_{j\neq\ i}^{n}{\frac{1}{4}d_{ij}\ x_j\ \left(1-x_i\right)}
                \end{equation}
                    \label{eq: expersion 19}
                given that $d_{ij}=d_{ji}$,
                \begin{equation}
                    \sum_{i}^{n-1}\sum_{j\neq\ i}^{n}{\frac{1}{4}d_{ij}\left(1-x_i\right)\left(1-x_j\right)}
                    =\sum_{i}^{n-1}\sum_{j\neq\ i}^{n}{\frac{1}{4}d_{ij}\left(1-x_i\right)}-\sum_{i}^{n-1}\sum_{j\neq\ i}^{n}{\frac{1}{4}d_{ij}\ x_i\ \left(1-x_j\right)}
                \end{equation}
                now, introducing this in Eq.~\ref{eq: expersion 19}:
                \begin{multline}
                        \sum_{i}^{n-1}\sum_{j>i}^{n}{\frac{1}{2}d_{ij}\left(1-x_ix_j\right)}\
                        \le\ \\ \le\ \sum_{i}^{n-1}\sum_{j\neq\ i}^{n}{\frac{1}{4}d_{ij}\left(1-x_i\right)}+\sum_{i}^{n-1}\sum_{j\neq\ i}^{n}{\frac{1}{4}d_{ij}x_i\ \left(1-x_j\right)}
                        +\sum_{i}^{n-1}\sum_{j\neq\ i}^{n}{\frac{1}{4}d_{ij}\left(1-x_i\right)}-\sum_{i}^{n-1}\sum_{j\neq\ i}^{n}{\frac{1}{4}d_{ij}\ x_i\ \left(1-x_j\right)}
                \end{multline}
                finally,
                \begin{equation}
                    \sum_{i}^{n-1}\sum_{j>i}^{n}{\frac{1}{2}d_{ij}\left(1-x_ix_j\right)}\ \le\ \ \sum_{i}^{n-1}\sum_{j\neq\ i}^{n}{\frac{1}{2}d_{ij}\left(1-x_i\right)}\
                \end{equation}
                The second term corresponds exactly to the sum of the weights of all edges incident to node $i$ for all variables satisfying $x_i = -1$, that is, for the $c$ nodes separated by the cut. When applied to the illustrative example discussed above, this expression recovers Eq.~(\ref{eq: example_cut_inequality}).

                We now introduce the quantity
                \begin{equation}
                d_i = \sum_{j=1}^{n} d_{ij},
                \end{equation}
                which represents the sum of all weights connected to node $i$. For a fixed value of $c$ in the constraint, let ${\bar{d}_i} | {i \in n_c}$ denote the set of the $c$ largest values of $d_i$, where $n_c$ indexes the corresponding nodes. It then follows that:
                \begin{equation}
                    \sum_{i}^{n-1}\sum_{j>i}^{n}{\frac{1}{2}d_{ij}\left(1-x_ix_j\right)}\ \le\ \ \sum_{i}^{n}{\frac{1}{2}d_i\left(1-x_i\right)}\le\sum_{i}^{n_c}{\bar{d}}_i
                \end{equation}
\newpage
\section{PB-PCE execution}

\begin{figure*}[htbp]
        \centering
        \includegraphics[width=\textwidth]{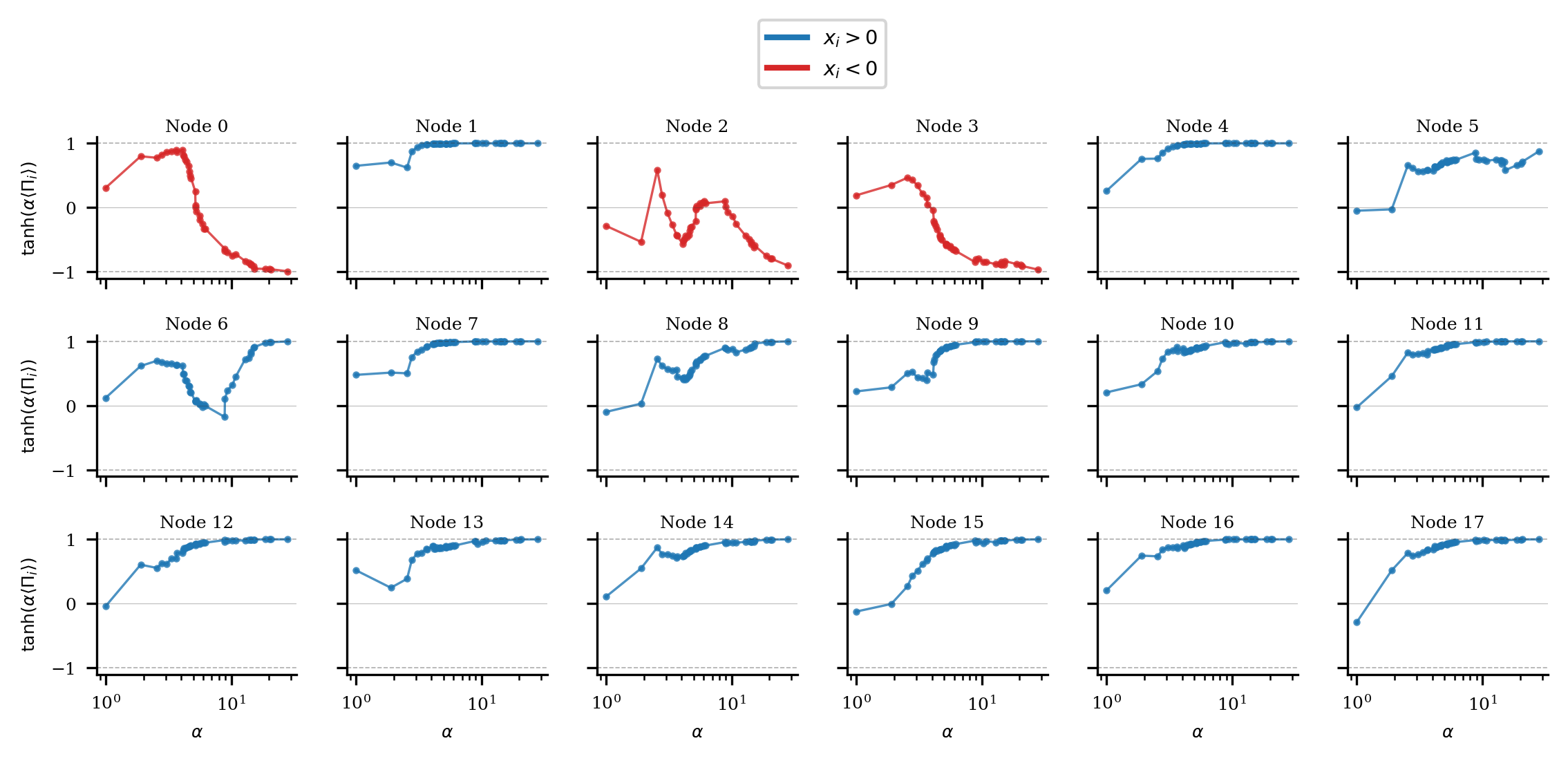}
        \caption{Evolution of the encoded variables during a single execution of PB-PCE. Each curve represents $\tanh(\alpha \langle \Pi_i \rangle)$ as a function of $\alpha$.}
        \label{fig: ejemplo de ejecucion iterativa}
    \end{figure*}

    In Figure~\ref{fig: ejemplo de ejecucion iterativa} we show the evolution of the variables throughout the full execution of PB-PCE. As $\alpha$ increases, most variables rapidly approach their binarized values and remain in the plateau once reached. In contrast, a small subset of variables continues to explore the loss landscape more deeply, exhibiting larger fluctuations until binarization is ultimately achieved.

\section{PB-PCE parameters}

    In Table~\ref{tab: especificaciones} we report the specific parameter used in all the PB-PCE executions. For the 150- and 300-node graph instances, we increased the constraint threshold. Due to the significantly larger number of possible variable combinations, we observed that some configurations with values as high as 0.90 were still able to violate the constraint. In addition, the initial value of the penalty parameter $\alpha$ was reduced to 1, as we found that using $\alpha=3$ caused the PCE to struggle to converge and to effectively minimize the constraint value itself, even when the constraint was not yet satisfied.
    
    \begin{table}[htbp]
        \centering
    \begin{tabular}{rrrrr}
    \toprule
    Nodes & Qubits & Order (k) & Threshold (M) & Initial $\alpha$ \\
    \midrule
    6   &  3 & 2 & 0.90                & 3                \\
    14  &  4 & 2 & 0.90                & 3                \\
    18  &  4 & 2 & 0.90                & 3                \\
    20  &  5 & 2 & 0.90                & 3                \\
    25  &  5 & 2 & 0.90                & 3                \\
    50  &  6 & 3 & 0.90                & 3                \\
    150 &  8 & 4 & 0.95                & 1                \\
    300 &  9 & 4 & 0.95                & 1                \\
    \bottomrule
    \end{tabular}
    \caption{Parameters used on the PB-PCE executions}
    \label{tab: especificaciones}
    \end{table}

\section{Comparison of the $\alpha$-update heuristics}
\label{appendix_heuristics}

\begin{figure*}[htbp]
            \centering
            \includegraphics[width=\textwidth]{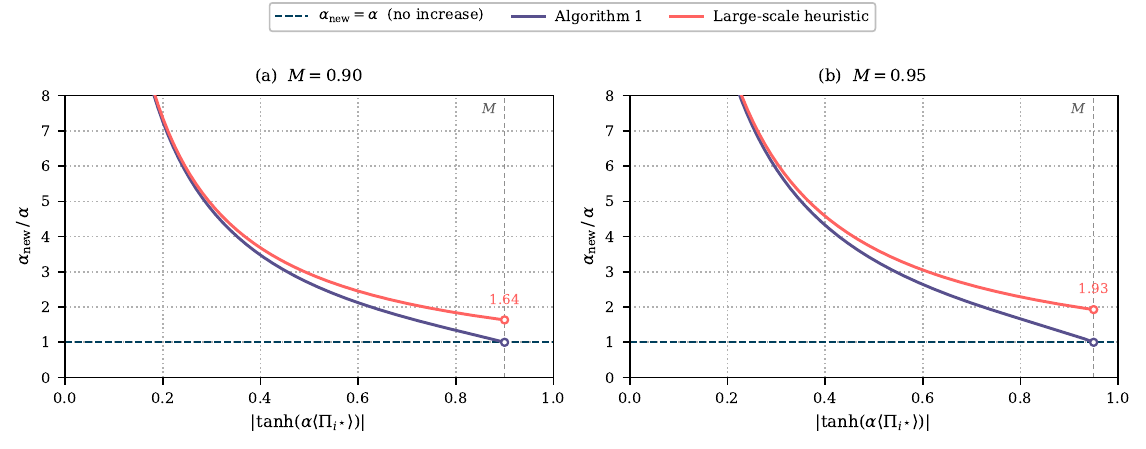}
            \caption{Multiplicative factor $\alpha_{\mathrm{new}}/\alpha$ applied by the two $\alpha$-update heuristics, as a function of the binarization value $\left|\tanh\left(\alpha\langle\Pi_{i^\star}\rangle\right)\right|$ of the selected variable, for (a) $M=0.90$ and (b) $M=0.95$. The open circles mark the limiting values reached at the threshold: the heuristic of Algorithm 1 degenerates to $\alpha_{\mathrm{new}}/\alpha \to 1$, i.e. no increase of $\alpha$, whereas the large-scale variant retains a strictly positive increment $\operatorname{arctanh}(M)/M$. Far from the threshold both curves coincide. The vertical axis is linear and clipped, so the divergence of both factors as the abscissa approaches zero runs off the top of the panels.}
            \label{fig: heuristics comparison}
        \end{figure*}

Both variants of the PB-PCE heuristic update the binarization parameter multiplicatively at the end of every stage, and differ only in the denominator of the applied factor. The rule used in Algorithm 1 is
\begin{equation}
    \alpha \gets \alpha
        \frac{\operatorname{arctanh}(M)}
             {\operatorname{arctanh}\left(\left|\tanh\left(\alpha\langle\Pi_{i^\star}\rangle\right)\right|\right)},
    \label{eq: heuristic algorithm 1}
\end{equation}
whereas the variant employed for the large-scale instances is
\begin{equation}
    \alpha \gets \alpha
        \frac{\operatorname{arctanh}(M)}
             {\left|\tanh\left(\alpha\langle\Pi_{i^\star}\rangle\right)\right|},
    \label{eq: heuristic large scale}
\end{equation}
where $i^\star$ denotes the least-binarized variable selected at that stage and $M$ is the binarization threshold. In both cases the applied factor depends on the circuit parameters only through the binarization value $\left|\tanh\left(\alpha\langle\Pi_{i^\star}\rangle\right)\right|$ of the selected variable, so the two rules can be compared directly as functions of this quantity, as shown in Figure~\ref{fig: heuristics comparison}.

The two rules differ precisely in the regime that caused the optimization to stall. When the selected variable already sits almost exactly at the threshold, $\left|\tanh\left(\alpha\langle\Pi_{i^\star}\rangle\right)\right| \to M$, the denominator of Eq.~(\ref{eq: heuristic algorithm 1}) tends to $\operatorname{arctanh}(M)$ and the whole factor tends to one, so that $\alpha$ is left essentially unchanged and the stage makes no progress. Under Eq.~(\ref{eq: heuristic large scale}) the factor instead tends to
\begin{equation}
    \frac{\operatorname{arctanh}(M)}{M} > 1 ,
\end{equation}
which is strictly larger than unity for every $M \in (0,1)$, since $\operatorname{arctanh}(M) > M$ in that interval. For the thresholds used in this work this limit equals $1.64$ for $M = 0.90$ and $1.93$ for $M = 0.95$, so an increase of $\alpha$ of at least that factor is guaranteed at every stage, no matter how close the selected variable is to the threshold.

In the opposite regime the two rules agree. When the selected variable is still far from the threshold, $\left|\tanh\left(\alpha\langle\Pi_{i^\star}\rangle\right)\right| \to 0$, and since $\operatorname{arctanh}(u) = u + \mathcal{O}(u^{3})$ the two denominators coincide to leading order, so both factors diverge in the same way. This is apparent in Figure~\ref{fig: heuristics comparison}, where the two curves are indistinguishable for small values of the abscissa and separate only as the threshold is approached. The modification therefore preserves the behaviour of the original heuristic while the variables are still far from being binarized, where large increments of $\alpha$ are required, and removes the vanishing step size that stalled the optimization close to the threshold.

\section{Large-scale analysis}
\label{appendix_large_scale}
\begin{figure*}[htbp]
            \centering
            \includegraphics[width=\textwidth]{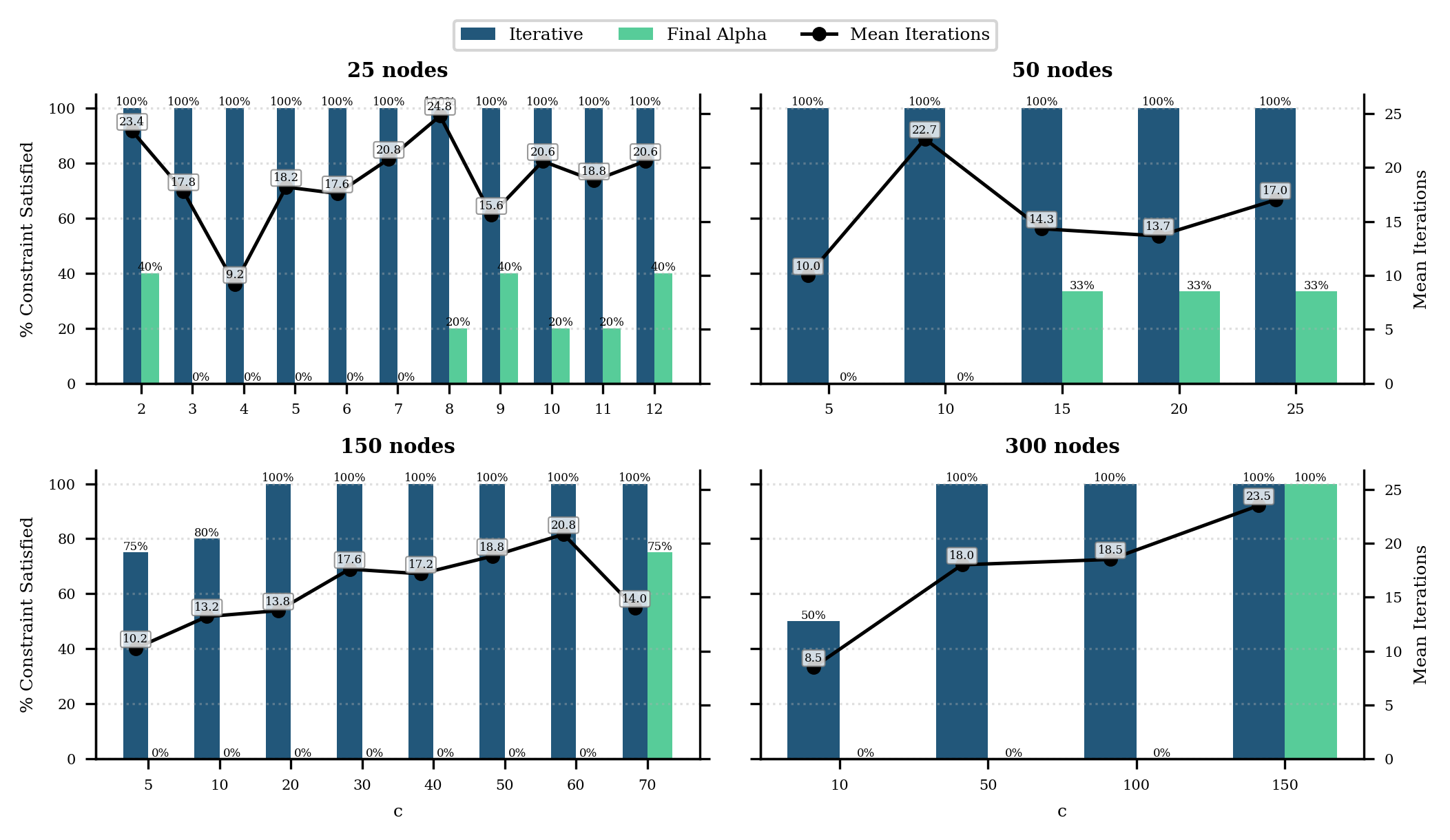}
            \caption{Performance comparison between PB-PCE method and the single execution of the PCE approach with the final $\alpha$ parameter found by the PB-PCE method for $n\in\{18, 50, 150, 300\}$ and different values of the parameter $c$ in the constraint term.}
            \label{fig_large_scale_results}
        \end{figure*}

Figure~\ref{fig_large_scale_results} shows a detailed performance comparison between the single PCE execution and the PB-PCE method for $n \in \{25, 50, 150, 300\}$ nodes. Each subplot reports the degree of constraint satisfaction for both approaches, for different values of the constraint parameter $c$, with single execution PCE shown in green and PB-PCE in blue. Results are consistent regardless the level of constraint imposed in each problem instance. PB-PCE consistently converges to higher-quality solutions, reaching full constraint satisfaction in most instances, while the single execution variant increasingly struggles as the problem size grows. These results underscore the effectiveness of the PB-PCE strategy for constrained optimization problems, particularly when scalability and solution reliability are essential.

\section{Runtime analysis}
\label{appendix_runtime}
Table~\ref{tab_runtime_comparison} compares PCE runtimes between the QARP framework (using Qulacs backend) and Qiskit \citep{javadi-abhari_quantum_2024}. Across all tested problem sizes, QARP consistently achieves lower execution times, while Qiskit exhibits a rapid increase as the number of nodes and qubits grows. This performance gap becomes more pronounced for larger instances.

    \begin{table}[htbp]
        \centering
        \begin{tabular}{rrcc}
        \toprule
        Nodes & Qubits & QARP (s) & Qiskit (s) \\
        \midrule
        6  & 3 & $5.52 \pm 0.42$   & $6.11 \pm 4.47$ \\
        18 & 4 & $11.24 \pm 0.81$  & $27.08 \pm 12.12$ \\
        25 & 5 & $17.44 \pm 1.26$  & $44.16 \pm 18.52$ \\
        50 & 7 & $39.29 \pm 3.98$  & $199.44 \pm 113.38$ \\
        \bottomrule
        \end{tabular}
        \caption{Runtime comparison between QARP and Qiskit.}
        \label{tab_runtime_comparison}
        \end{table}

\end{document}